\documentclass[journal, twoside]{IEEEtran}
\usepackage{amsmath,amsfonts}
\usepackage{algorithmic}
\usepackage{algorithm}
\usepackage{array}
\usepackage[caption=false,font=footnotesize,labelfont=rm,textfont=sf]{subfig}
\usepackage{textcomp}
\usepackage{stfloats}
\usepackage{url}
\usepackage{verbatim}
\usepackage{graphicx}
\usepackage{cite}
\usepackage{color}
\usepackage[strings]{underscore}
\usepackage{makecell}
\usepackage{bm}

\begin{document}

\title{High-Speed Ultra-Energy-Efficient Memristor-Based Massive MIMO SIC Detector Circuit With Hybrid Analog-Digital Computing Architecture}

\author{Jia-Hui~Bi, Shaoshi~Yang,~\IEEEmembership{Senior Member,~IEEE}, Sheng~Chen,~\IEEEmembership{Life~Fellow,~IEEE}, and Ping~Zhang,~\IEEEmembership{Fellow,~IEEE} 
\vspace{-1.5mm}

\thanks{Copyright (c) 2025 IEEE. Personal use of this material is permitted. However, permission to use this material for any other purposes must be obtained from the IEEE by sending a request to pubs-permissions@ieee.org.}
\thanks{This work was supported by Beijing Municipal Natural Science Foundation under Grant L242013. \textit{(Corresponding author: Shaoshi Yang.)}} %
\thanks{J.-H.~Bi, S.~Yang and P.~Zhang are with the School of Information and Communication Engineering, Beijing University of Posts and Telecommunications, Beijing 100876, China (e-mail: bijiahui@bupt.edu.cn, shaoshi.yang@bupt.edu.cn, pzhang@bupt.edu.cn).}
\thanks{S. Chen is with the School of Electronics and Computer Science, University of Southampton, SO17 1BJ Southampton, U.K. (e-mail: sqc@ecs.soton.ac.uk).}
\thanks{Digital Object Identifier 10.1109/TVT.2025.3544093}
}

\markboth{Published in IEEE Transactions on Vehicular Technology}{Bi \MakeLowercase{\textit{et al.}}: High-Speed Ultra-Energy-Efficient Memristor-Based Massive MIMO SIC Detector Circuit}

\maketitle

\begin{abstract}
The emerging memristor crossbar array based computing circuits exhibit computing speeds and energy efficiency far surpassing those of traditional digital processors. This type of circuits can complete high-dimensional matrix operations in an extremely short time through analog computing, making it naturally applicable to linear detection and maximum likelihood detection in massive multiple-input multiple-output (MIMO) systems. However, the challenge of employing memristor crossbar arrays to efficiently implement other nonlinear detection algorithms, such as the successive interference cancellation (SIC) algorithm, remains unresolved. In this paper we propose a memristor-based circuit design for massive MIMO SIC detector. The proposed circuit comprises several judiciously designed analog matrix computing modules and hybrid analog-digital slicers, which enables the proposed circuit to perform the SIC algorithm with a hybrid analog-digital computing architecture. We show that the computing speed and the computational energy-efficiency of the proposed detector circuit are 43 times faster and 110 times higher, respectively, than those of a traditional 8-core digital signal processor (DSP), and also advantageous over the benchmark high-performance field programmable gate array (FPGA) and graphics processing unit (GPU).
\end{abstract}


\section{Introduction}\label{S1}

\IEEEPARstart{I}{n} modern wireless communication systems, the massive multiple-input multiple-output (MIMO) technology is a key enabler, which employs a large number of antennas to simultaneously serve multiple users, thus significantly enhancing the transmission rates and spectral efficiency. However, the extensive use of radio frequency (RF) chains in massive MIMO incurs substantial power consumption. Additionally, the large number of antennas significantly increases the complexity of baseband signal processing algorithms, such as signal detection algorithms \cite{FiftyYears}, resulting in high processing latency and energy consumption in the baseband. Therefore, designing low-latency, energy-efficient massive MIMO communication systems has long been a research hotspot in the field of wireless communications. To address this challenge, extensive research efforts have been undertaken. For example, the hybrid analog/digital architecture has been proposed as an effective means of reducing the number of RF chains, yielding a valuable solution for enhancing the energy efficiency of massive MIMO systems \cite{HAD1,HAD2}. On the other avenue, the emerging memristor devices have recently gained significant attention due to their great potential in realizing low-latency, high-energy-efficiency massive MIMO baseband signal processors.

Memristors are typically integrated into crossbar arrays, which enables high-dimensional matrix operations, such as matrix multiplication and inversion \cite{SunZhong_Inv}, to be completed within an extremely short time, generally in the range of tens of nanoseconds. The underlying principle of memristor-based matrix computing involves mapping the matrix operand onto the conductance matrix of the memristor crossbar array, mapping the vector operand onto the input voltages or currents, and obtaining computational results by measuring the output voltages of the circuit \cite{MemristorSurvey}. This form of analog matrix computing is an in-memory computing approach, which bypasses the von Neumann bottleneck, thereby achieving significantly higher processing speed and computational energy efficiency than the traditional digital computing approach.

Memristor-based analog matrix computing technology was initially used to accelerate matrix multiplications in neural network training and has been applied to massive MIMO baseband signal processing, especially in MIMO detection, in recent years. The work \cite{MemristorBaseband} first applied memristor-based analog matrix computing technology to MIMO signal detection, by utilizing memristor crossbar arrays to perform the matrix multiplication operations in the minimum mean square error (MMSE) detection. The study \cite{RidgeRegression} proposed a memristor-based ridge regression computing circuit and applied it to perform linear detection algorithms, and the study \cite{InMemoryBaseband} introduced a similarly structured memristor-based linear detector circuit. The work \cite{RethinkingDetection} proposed a memristor-based MMSE detection scheme by converting the MMSE algorithm into a linearized iterative algorithm and using memristor crossbar arrays to accelerate it. The work \cite{YiHangRen} employed memristor crossbar arrays to perform the matrix multiplication operations in the maximum likelihood (ML) detection with ultra-high energy efficiency. The superior performances of memristor-based massive MIMO detectors have been demonstrated in the aforementioned works.

Existing works utilizing memristor crossbar arrays for analog matrix computing can naturally achieve linear detection and ML detection, because the processes of linear and ML detection algorithms primarily involve matrix computations. However, the processes of other nonlinear detection algorithms are often more complex and do not rely solely on matrix computations. For instance, the successive interference cancellation (SIC) detection, a widely used detection algorithm with better detection performance than linear detection, involves iterative matrix computations and slicing operations. Existing memristor-based computing circuits are limited to performing matrix computations but cannot perform slicing operations. Therefore, efficiently performing the SIC algorithm using memristor crossbar arrays remains an unresolved challenge.

In this paper, we present a memristor-based massive MIMO SIC detector circuit. The proposed detector circuit comprises several memristor-based matrix computing modules and associated slicers. The proposed matrix computing modules perform the matrix computations in the SIC algorithm through analog computing. The proposed hybrid analog-digital slicers first convert each of the analog voltages to be quantized into a digital (binary) vector and then output the corresponding quantization voltage levels based on these binary vectors. Therefore, the proposed detector circuit implements the SIC algorithm through a hybrid analog-digital processing approach. We investigate the impact of the precision of memristors on detection performance of the proposed detector. We also evaluate the computing speed and computational energy efficiency of the proposed circuit. Our results show that they are 43 times faster and 110 times higher, respectively, than those of a traditional 8-core digital signal processor (DSP) in our considered scenario, and also advantageous over the benchmark high-performance field programmable gate array (FPGA) and graphics processing unit (GPU).

\section{MIMO System and SIC Detection}\label{S2}

\subsection{System Model}\label{S2.1}

We consider a massive MIMO system, in which the base station (BS) is equipped with $R$ antennas to support $K$ single-antenna user terminals (UTs) with $R>K$. Let $\lambda_1, \lambda_2, \cdots , \lambda_K$ be the transmitting power values of UTs, respectively. The uplink received signals are given by:
\begin{equation}\label{y=HPs+n_Complex} 
	{\mathbf{y}} = {\mathbf{H}} \mathbf{\Lambda} {\mathbf{s}} + {\mathbf{n}}, 
\end{equation}
where ${\mathbf{y}}\! \in\! \mathbb{C}^{R\times 1}$ is the received signal vector, ${\mathbf{s}}\! \in\! \mathbb{C}^{K\times 1}$ is the transmitted signal vector sent by the UTs, ${\mathbf{H}}\! \in\! \mathbb{C}^{R\times K}$ is the channel matrix, the diagonal matrix ${\mathbf{\Lambda}}\! =\! \textrm{diag}\big(\sqrt{\lambda_1}, \sqrt{\lambda_2}, \cdots , \sqrt{\lambda_K}\big)$, and ${\mathbf{n}}\! \in\! \mathbb{C}{^{R\times 1}}$ is a zero-mean complex additive white Gaussian noise (AWGN) vector having the covariance matrix $\mathbb{E}\left[{\mathbf n} {\mathbf n} ^{\textrm {H}}\right] = \sigma_n^2 \mathbf{I}$, with $(\cdot )^{\textrm {H}}$ denoting the Hermitian transpose operator and $\mathbf{I}$ denoting the identity matrix of appropriate dimension. The transmitted signals are normalized as $\mathbb{E} \left[ \left| s_k \right| {^2} \right]\! =\! 1$, $1\le k\le K$, where $s_k$ is the $k$th element of ${\mathbf s}$. By defining ${\mathbf{F}}\! =\! {\mathbf{H}}\mathbf{\Lambda}\! \in\! \mathbb{C}^{R\times K}$, we obtain:
\begin{equation}\label{y=Fs+n_Real} 
	{\mathbf{y}} = {\mathbf{F}} {\mathbf{s}} + {\mathbf{n}}.
\end{equation}

\subsection{Operations of SIC Detection}\label{S2.2}

The core idea of the SIC detection algorithm is to detect the transmitted symbols one by one using a linear detection algorithm, with the aid of cancelling the corresponding detected symbols from the received signals after each detection \cite{SIC,MIMOOFDM}. We consider the MMSE-SIC algorithm in this paper.

Let $\left\{m_1, m_2, \cdots ,m_K\right\}$ be the ordered set representing the detection sequence, and ${\mathbf{G}}$ be the reordered transfer matrix, i.e., ${\mathbf{G}}\! =\! [{\mathbf f}_{m_1}, {\mathbf f}_{m_2}, \cdots, {\mathbf f}_{m_K}]$ with ${\mathbf f}_k$ being the $k$th column of ${\mathbf F}$. Denote by ${\mathbf G}_{(k)}$ the first to $k$th columns of ${\mathbf G}$, and by ${\mathbf G}_{\langle k \rangle }$ the $k$th to $K$th columns of ${\mathbf G}$, i.e., ${\mathbf G}_{(k)} \! =\! [{\mathbf f}_{m_1}, \cdots, {\mathbf f}_{m_k}]$, ${\mathbf G}_{\langle k \rangle}\! =\! [{\mathbf f}_{m_k}, \cdots, {\mathbf f}_{m_K}]$. Let $e_{m_k}$ be the $k$th estimated symbol, i.e., the estimated value of $s_{m_k}$. Denote by ${\mathbf e}_{(k)}$ the vector consisting of the first to $k$th estimated symbols, i.e., ${\mathbf e}_{(k)}\! =\! [e_{m_1}, \cdots , e_{m_k}]^{\textrm {T}}$, where $(\cdot )^{\textrm {T}}$ denotes the transpose operator.

The SIC algorithm performs the matrix computation
\begin{equation}\label{SICcomplex1} 
	{\mathbf b}_1 = \big({\mathbf{G}}^{\textrm {H}} {\mathbf{G}} + \sigma_n^2 \mathbf{I} \big)^{-1} {\mathbf{G}}^{\textrm {H}} {\mathbf{y}}
\end{equation}
to estimate $e_{m_1}$, where ${\mathbf b}_1$ is the result vector, and $(\cdot )^{-1}$ denotes the inverse operator. Specifically, $e_{m_1}$ is obtained by performing a slicing (i.e., quantization) operation on the first element of ${\mathbf b}_1$.

The SIC algorithm performs the matrix computation
\begin{equation}\label{SICcomplexk} 
	{\mathbf b}_k = \big({\mathbf{G}}_{\langle k \rangle }^{\textrm {H}} {\mathbf{G}}_{\langle k \rangle } + \sigma_n^2 \mathbf{I} \big)^{-1} {\mathbf{G}}_{\langle k \rangle }^{\textrm {H}} \big({\mathbf{y}} - {\mathbf{G}}_{(k-1)} {\mathbf{e}}_{(k-1)}\big)
\end{equation}
to estimate $e_{m_k}(1< k\leq K)$, where ${\mathbf b}_k$ is the result vector. Similarly, $e_{m_k}$ is obtained by performing a slicing operation on the first element of ${\mathbf b}_k$.

Obviously, the SIC detection processes consist of $K$ matrix computations, and each of them is followed by a slicing operation. Since the memristor-based matrix computing circuit can only perform real-valued matrix computations, we define two conversion operators $\mathcal{V} (\cdot )$ and $\mathcal{M} (\cdot )$ for converting a complex-valued vector to a real-valued vector and a complex-valued matrix to a real-valued matrix, respectively:
\begin{align*}
  & \mathcal{V}\big(\cdot \big) = \left[ \begin{array}{c}
  \Re \big( \cdot \big) \\
  \Im \big( \cdot \big)
  \end{array} \right] , ~
  \mathcal{M}\big( \cdot \big) = \left[ \begin{array}{cc}
  \Re \big( \cdot \big) & -\Im \big( \cdot \big) \\
  \Im \big( \cdot \big) &  \Re \big( \cdot \big)
  \end{array} \right],
\end{align*} 
where $\Re (\cdot)$ and  $\Im (\cdot)$ denote the real and imaginary parts of the corresponding vector or matrix, respectively.

Let ${\mathbf r}_k$, satisfying ${\mathbf{r}}_k = \mathcal{V}({\mathbf b}_k)$, be the real-valued result vector of the $k$th matrix computation. Then the first matrix computation in SIC detection, i.e., \eqref{SICcomplex1}, can be expressed as:
\begin{equation}\label{MMSE1}
	{\mathbf r}_1 = \big(\tilde{\mathbf{G}}^{\textrm {T}} \tilde{\mathbf{G}} + \sigma_n^2 \mathbf{I} \big)^{-1} \tilde{\mathbf{G}}^{\textrm {T}} \tilde{\mathbf{y}},
\end{equation}
where $\tilde{\mathbf{G}} = \mathcal{M}({\mathbf{G}})$, $\tilde{\mathbf{y}} = \mathcal{V}({\mathbf{y}})$. The $k$th $(1< k\leq K)$ matrix computation, i.e., \eqref{SICcomplexk}, can be equivalently expressed as:
\begin{equation}\label{MMSEk}
	{\mathbf r}_k = \big(\tilde{\mathbf{G}}_{\langle k \rangle }^{\textrm {T}} \tilde{\mathbf{G}}_{\langle k \rangle } + \sigma_n^2 \mathbf{I} \big)^{-1} \tilde{\mathbf{G}}_{\langle k \rangle }^{\textrm {T}} \big(\tilde{\mathbf{y}} - \tilde{\mathbf{G}}_{(k-1)} \tilde{\mathbf{e}}_{(k-1)}\big),
\end{equation}
where $\tilde{\mathbf{G}}_{\langle k \rangle } = \mathcal{M}({\mathbf{G}}_{\langle k \rangle })$, $\tilde{\mathbf{G}}_{( k-1 )} = \mathcal{M}({\mathbf{G}}_{(k-1)})$, $\tilde{\mathbf{e}}_{(k-1)}=\mathcal{V}({\mathbf e}_{(k-1)})$.

Let $\mathcal{Q} (a_1, a_2)$ be the slicing operator, with the real values $a_1$ and $a_2$ corresponding to the real and imaginary parts of the sliced complex value, respectively. Denote by $r_{k}(k')$ the $k'$th element of ${\mathbf r}_k$. The $k$th $(1\leq k\leq K)$ slicing operation can be expressed as:
\begin{equation}\label{Slice}
	{e}_{m_k} = \mathcal{Q} \big(r_{k}(1), r_{k}(K+2-k)\big).
\end{equation}

\section{Proposed SIC Detector Circuit Design}\label{S3}

\subsection{Circuit Structure}\label{S3.1}

\begin{figure}[tbp]
  \centerline{\includegraphics[width=0.88\linewidth]{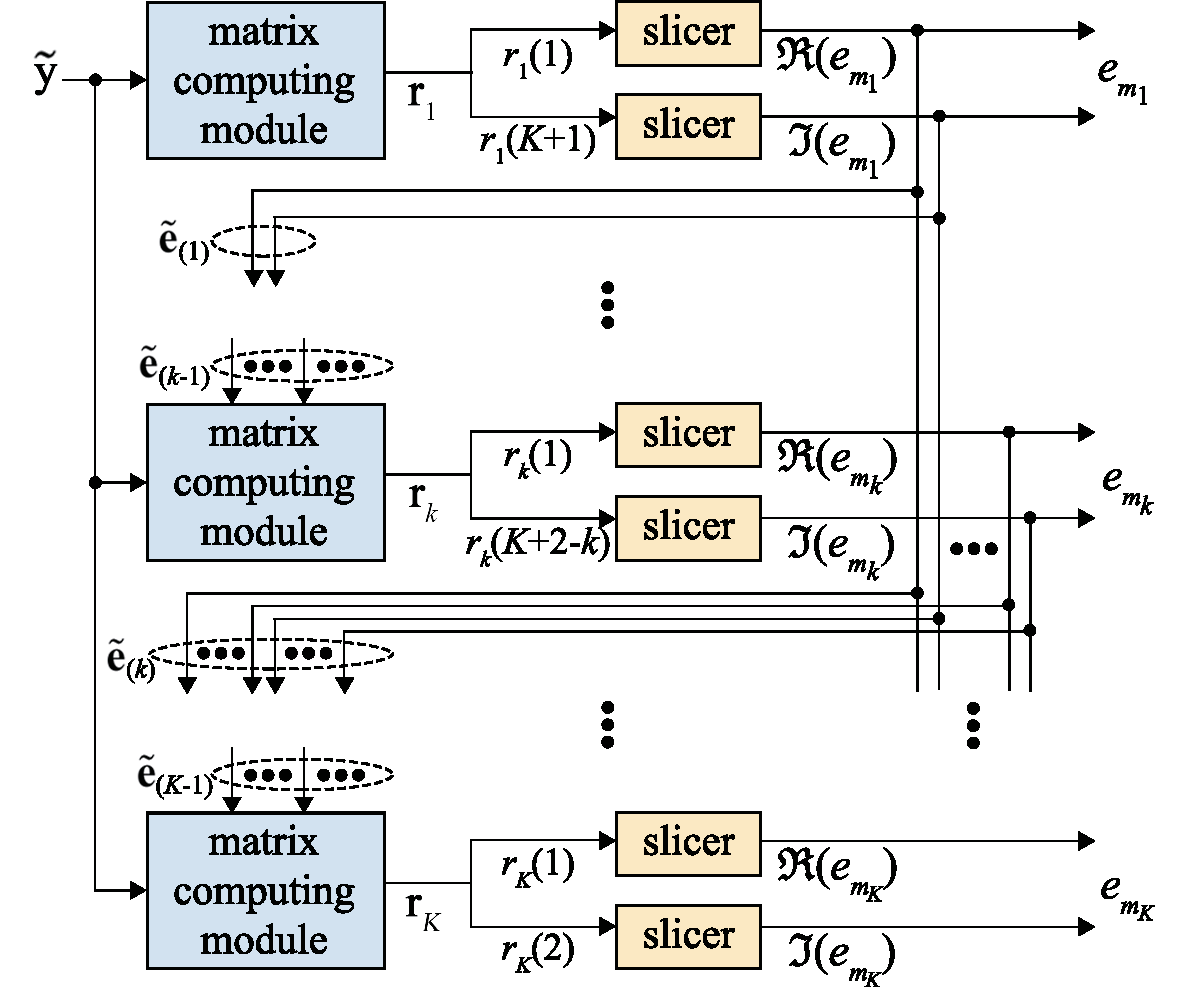}}
  \vspace*{-3.5mm}
  \caption{Structure of the proposed memristor-based SIC detector circuit.}
  \vspace*{-3mm}
  \label{detector} 
\end{figure}

Structure of the proposed SIC detector circuit is illustrated in Fig.~\ref{detector}, which consists of $K$ stages, each comprising a matrix computing module and two slicers. The matrix computing modules are employed to compute \eqref{MMSE1} and \eqref{MMSEk}, while the slicers are employed to perform \eqref{Slice}. In the $k$th stage, the matrix computing module outputs ${\mathbf r}_k$, where $r_k(1)$ and $r_k(K+2-k)$ serve as inputs to two slicers. The two slicers output the real and imaginary parts of $e_{m_k}$, respectively. The outputs of the $k$th stage serve as inputs to the $(k+1)$th to $K$th matrix computing modules, with $\tilde{\mathbf y}$ simultaneously serving as an input to each matrix computing module.

\subsection{Proposed Matrix Computing Module}\label{S3.2}

\begin{figure}[bp]
\vspace*{-4mm}
  \centerline{\includegraphics[width=0.9\linewidth]{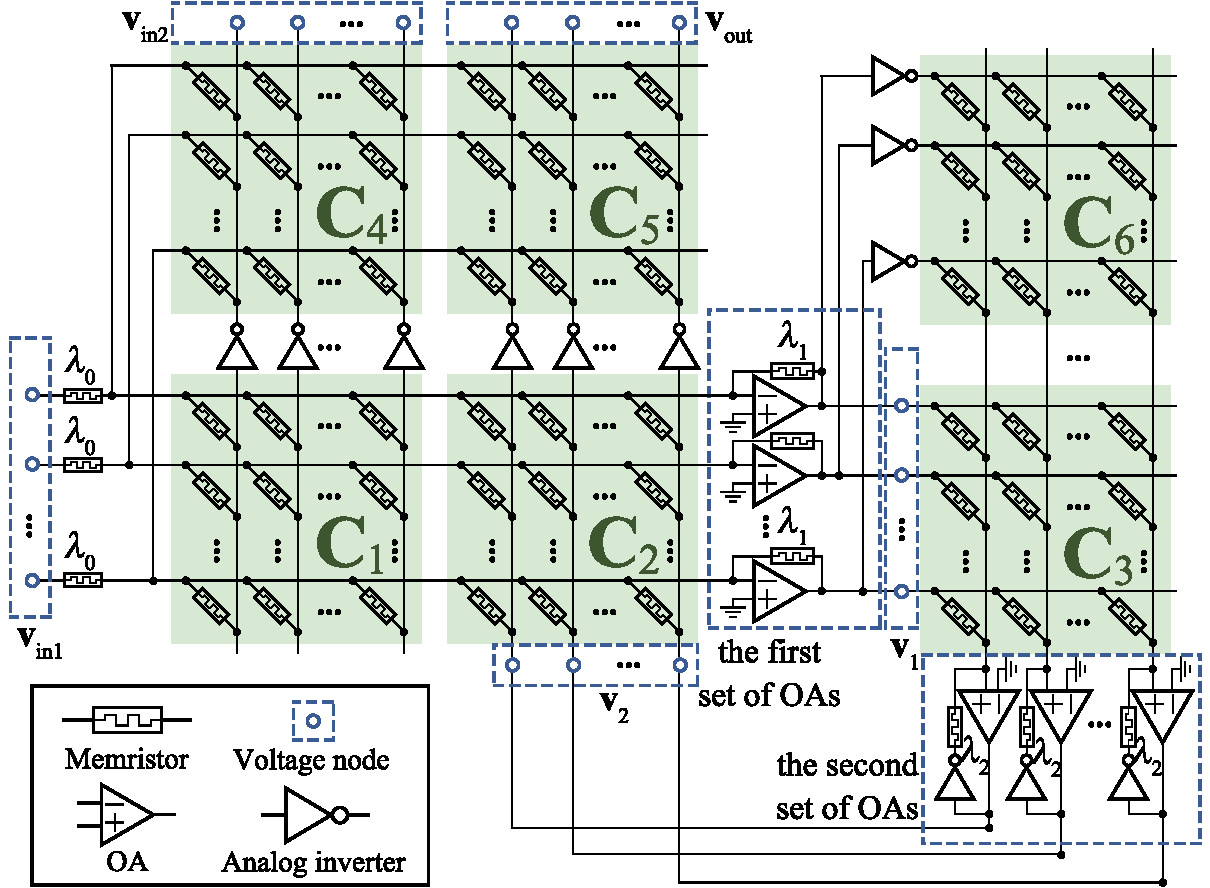}}
  \vspace*{-1mm}
  \caption{Circuit design of the proposed matrix computing module.}
  \vspace*{-0.5mm}
  \label{MCM} 
\end{figure}

The circuit design of the proposed matrix computing module is illustrated in Fig.~\ref{MCM}, which comprises six memristor crossbar arrays, three sets of analog inverters and two sets of operational amplifiers (OAs).

Let ${\mathbf C}_1$, ${\mathbf C}_2$, ${\mathbf C}_3$, ${\mathbf C}_4$, ${\mathbf C}_5$ and ${\mathbf C}_6$ be the conductance matrices of the six arrays, respectively, ${\mathbf v}_{\textrm {in1}}$ and ${\mathbf v}_{\textrm {in2}}$ be the two sets of input voltages, and ${\mathbf v}_{\textrm {out}}$ be the output voltages of the circuit, as shown in Fig.~\ref{MCM}. Let ${\mathbf v}_1$ and ${\mathbf v}_2$ be the voltages at the output nodes of the two sets of OAs, respectively. Clearly, ${\mathbf v}_{\textrm {out}}\! =\! -{\mathbf v}_2$. The conductance values of the memristors that are connected to the voltage nodes of ${\mathbf v}_{\textrm {in1}}$, the feedback memristors of the first set of OAs, and the feedback memristors of the second set of OAs are $\lambda_0$, $\lambda_1$ and $\lambda_2$, respectively.

The high gain of an OA results in a minimal voltage difference between its inverting and noninverting input nodes, making the voltages at the inverting input nodes of the first set of OAs and the noninverting input nodes of the second set of OAs approximate zeros. Additionally, the inherent properties of OAs result in approximate zero currents flowing into their inverting and noninverting input nodes. Based on Ohm’s law and Kirchhoff’s current law, the voltages in Fig.~\ref{MCM} satisfy:
\begin{equation}\label{Voltage1} 
	\lambda_0 {\mathbf v}_{\textrm {in1}} + ({\mathbf C}_4 - {\mathbf C}_1){\mathbf v}_{\textrm {in2}} + ({\mathbf C}_2 - {\mathbf C}_5){\mathbf v}_2 + \lambda_1 {\mathbf v}_1 = {\mathbf 0},
\end{equation}
and
\begin{equation}\label{Voltage2} 
	({\mathbf C}_3 - {\mathbf C}_6)^{\textrm {T}}{\mathbf v}_1 - \lambda_2 {\mathbf v}_2 = {\mathbf 0}.
\end{equation}
Upon substituting \eqref{Voltage2} into \eqref{Voltage1} we obtain:
\begin{equation}\label{Voltage3} 
	{\mathbf v}_{\textrm {out}} = ({\mathbf D}_3^{\textrm {T}} {\mathbf D}_2 + \lambda_1 \lambda_2 {\mathbf I})^{-1} {\mathbf D}_3^{\textrm {T}} (\lambda_0 {\mathbf v}_{\textrm {in1}} - {\mathbf D}_1{\mathbf v}_{\textrm {in2}}),
\end{equation}
where ${\mathbf D}_1 = {\mathbf C}_1 - {\mathbf C}_4$, ${\mathbf D}_2 = {\mathbf C}_2 - {\mathbf C}_5$, ${\mathbf D}_3 = {\mathbf C}_3 - {\mathbf C}_6$.

The conductance value of a memristor can be adjusted by a specific program \cite{SunZhong_Inv} to any value within a certain range. For the first stage, by mapping $\tilde{\mathbf y}$ onto $\lambda_0{\mathbf v}_{\textrm {in1}}$, mapping $\tilde{\mathbf G}$ onto ${\mathbf D}_2$ and ${\mathbf D}_3$, mapping $\sigma_n^2$ onto $\lambda_1 \lambda_2$, removing ${\mathbf C}_1$ and ${\mathbf C}_4$, the result of \eqref{MMSE1}, i.e., ${\mathbf r}_1$, can be obtained by measuring ${\mathbf v}_{\textrm {out}}$. For the $k$th $(1< k\leq K)$ stage, by mapping $\tilde{\mathbf y}$ onto $\lambda_0{\mathbf v}_{\textrm {in1}}$, mapping $\tilde{\mathbf G}_{(k-1)}$ onto ${\mathbf D}_1$, mapping $\tilde{\mathbf e}_{(k-1)}$ onto ${\mathbf v}_{\textrm {in2}}$, mapping $\tilde{\mathbf G}_{\left\langle k\right\rangle }$ onto ${\mathbf D}_2$ and ${\mathbf D}_3$, and mapping $\sigma_n^2$ onto $\lambda_1 \lambda_2$, the result of \eqref{MMSEk}, i.e., ${\mathbf r}_k$, can be obtained by measuring ${\mathbf v}_{\textrm {out}}$. Since the mapped matrix may contain both positive and negative elements, we map the computed matrix onto the difference between two positive conductance matrices.

Unlike traditional processors based on digital computing approach, the proposed matrix computing module employs the aforementioned analog computing approach to perform matrix computations in the SIC algorithm, thereby avoiding excessive time and computational resource consumption typically incurred by digital processor based matrix inversion operations.

\subsection{Proposed Hybrid Analog-Digital Slicer}\label{S3.3}

Let $\mathcal{S}_{\textrm {value}}\! =\! \{x_1, \cdots, x_W\}$ be the set of voltage values corresponding to all the possible values of $\Re(e_{m_k})$ or $\Im(e_{m_k})$, where $W$ represents the number of elements in the set and it depends on the modulation scheme. Let $\mathcal{S}_{\textrm {threshold}}\! =\! \{z_1, \cdots, z_{W-1}\}$ be the threshold set, i.e., $z_w\! =\! \frac{x_w + x_{w+1}}{2}$ $(1\leq w<W)$. The function of the proposed slicer is to slice a voltage into the element of $\mathcal{S}_{\textrm {value}}$ that is the closest to it. We propose two slicer structures, one termed the directly select structure, and the other termed the indirectly select structure.

\begin{figure}[tbp]
\vspace*{-1mm}
  \centering
    \subfloat[]{\includegraphics[width=\linewidth]{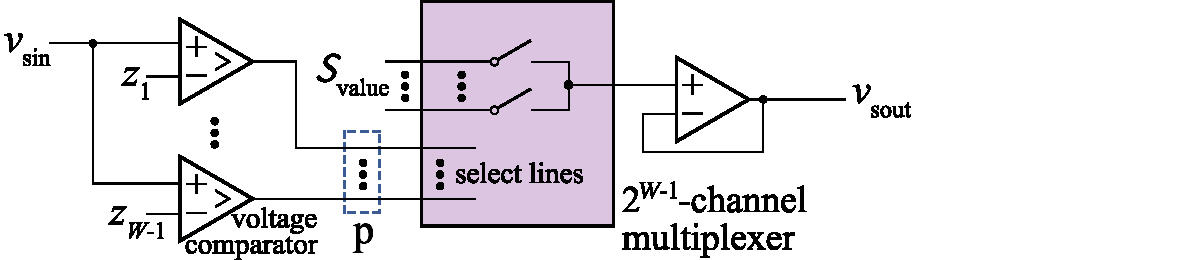}}\vspace*{-2mm}
    \hfil
    \subfloat[]{\includegraphics[width=\linewidth]{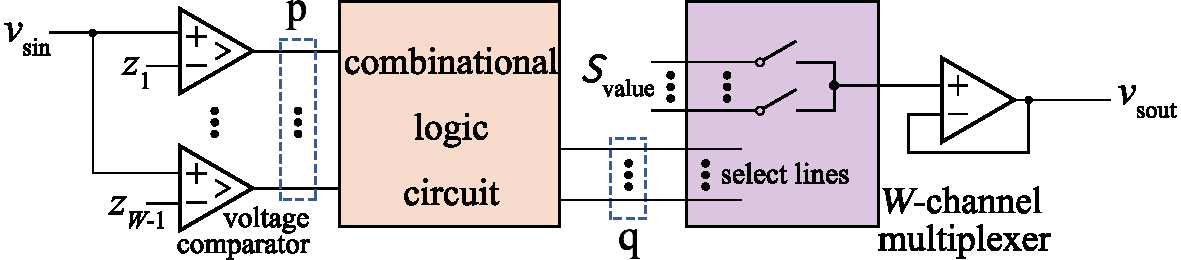}}\vspace*{-1mm}
  \caption{Structures of the proposed slicer: (a) the directly select structure, and (b) the indirectly select structure.}
\vspace*{-5mm}
\label{slicerstructure} 
\end{figure}

The directly select structure is illustrated in Fig.~\ref{slicerstructure}\,(a), which consists of $W\! -\! 1$ voltage comparators, a $2^{W\! -\! 1}$-channel analog multiplexer and an OA. When the voltage at the noninverting input node of a comparator is greater than that at its inverting input node, it outputs a high level, otherwise it outputs a low level. Let $v_{\textrm {sin}}$ and $v_{\textrm {sout}}$ be the input and output voltage of the proposed slicer, respectively. The comparators compare $v_{\textrm {sin}}$ with all the threshold voltages. Let $\mathbf p$ be the binary vector of length $W\! -\! 1$ formed by the outputs of the comparators, which are also employed as the select lines of multiplexer. Voltages with values of $\mathcal{S}_{\textrm {value}}$ serve as inputs to the appropriate input channels of multiplexer. The different magnitude relationships between $v_{\textrm {sin}}$ and threshold voltages result in different $\mathbf p$, causing the multiplexer to select different channels and output the corresponding values of $\mathcal{S}_{\textrm {value}}$. The OA is employed to construct a voltage follower to ensure output voltage stability.

The indirectly select structure is illustrated in Fig.~\ref{slicerstructure}\,(b), which consists of $W\! -\!1$ voltage comparators, a combinational logic circuit, a $W$-channel analog multiplexer and an OA. The combinational logic circuit transforms $\mathbf p$ into a shorter binary vector $\mathbf q$ of length $\log_2 W$, so as to indirectly select a channel of the multiplexer. The indirectly select structure can enhance the utilization of multiplexer channels, especially when the modulation order is high. For instance, when utilizing 64 quadrature amplitude modulation (QAM), i.e., $W\! =\! 8$, the directly select structure necessitates a $128$-channel analog multiplexer, whereas the indirectly select structure only requires an $8$-channel multiplexer.

\begin{figure}[bp!]
  \vspace*{-4mm}
  \centering
    \subfloat[]{\includegraphics[width=\linewidth]{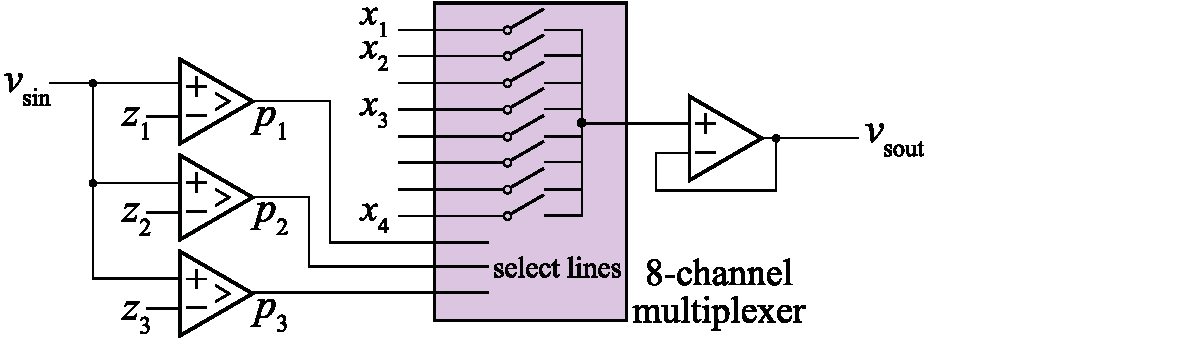}}\vspace*{-4mm}
    \hfil
    \subfloat[]{\includegraphics[width=\linewidth]{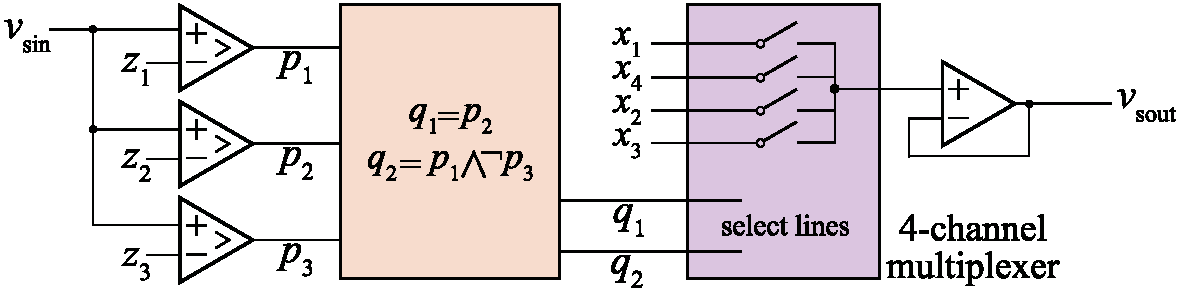}}\vspace*{-1mm}
  \caption{The proposed slicer with: (a) the directly select structure, and (b) the indirectly select structure, using 16 QAM as an example.}
  \vspace*{-0.5mm}
  \label{slicer} 
\end{figure}

To provide a clearer illustration of the two structures, we consider the 16 QAM example. Hence, $x_1\! =\! - \frac{3}{{\sqrt{10}}}v_0$, $x_2\! =\! - \frac{1}{{\sqrt{10}}}v_0$, $x_3\! =\! \frac{1}{{\sqrt{10}}}v_0$ and $x_4\! =\!\frac{3}{{\sqrt{10}}}v_0$, while $z_1\! =\! - \frac{2}{{\sqrt{10}}}v_0$, $z_2\! =\! 0$ and $z_3\! =\! \frac{2}{{\sqrt{10}}}v_0$, where $v_0$ is a reference voltage. The varying values of $v_{\textrm {sin}}$ lead to the four possible values of $\mathbf p$: $[0,0,0]$, $[1,0,0]$, $[1,1,0]$, $[1,1,1]$. The slicer with the directly select structure is illustrated in Fig.~\ref{slicer}\,(a), the four possible $\mathbf p$ values select the first, second, fourth and eighth input channels of the 8-channel multiplexer, respectively. So we input $x_1$, $x_2$, $x_3$ and $x_4$ to the corresponding four channels. The slicer with the indirectly select structure is illustrated in Fig.~\ref{slicer}\,(b). The logic formulas of the combinational logic circuit are $q_1\! =\! p_2$ and $q_2\! =\! p_1 \wedge\neg  p_3$, where $\neg$ and $\wedge$ denote the logical NOT and AND, respectively, and it is worth noting that this is not the only option. The four values of $\mathbf q$ corresponding to the four possible values of $\mathbf p$ are: $[0,0]$, $[0,1]$, $[1,1]$, $[1,0]$, which select the first, third, fourth, second input channels of the 4-channel multiplexer, respectively. So we input $x_1$, $x_2$, $x_3$ and $x_4$ to the four channels. Table~\ref{relationship} summarizes the relationship between $v_{\textrm {sin}}$, $\mathbf p$, $\mathbf q$ (if exist) and $v_{\textrm {sout}}$ in both the directly select structure and indirectly select structure.

\renewcommand{\arraystretch}{1.25}
\begin{table}[!t]
\caption{Relationship between $v_{\textrm sin}$, $\mathbf p$, $\mathbf q$, and $v_{\textrm sout}$ for $W=4$}
\label{relationship} 
\vspace*{-2mm}
\centering
  \begin{tabular}{|>{\centering}p{0.2\linewidth}|>{\centering}p{0.2\linewidth}|>{\centering}p{0.2\linewidth}|p{0.1\linewidth}<{\centering}|}
  \hline
  $v_{\textrm {sin}}$ & $\mathbf p$ & $\mathbf q$ & $v_{\textrm {sout}}$ \\
  \hline
  $<z_1$ & $[0,0,0]$ & $[0,0]$ & $x_1$ \\
  \hline
  $z_1\sim z_2$ & $[1,0,0]$ & $[0,1]$ & $x_2$ \\
  \hline
  $z_2\sim z_3$ & $[1,1,0]$ & $[1,1]$ & $x_3$ \\
  \hline
  $>z_3$ & $[1,1,1]$ & $[1,0]$ & $x_4$ \\
  \hline
  \end{tabular}
\vspace*{-3mm}
\end{table}

\section{Simulations}\label{S4}

Let $\alpha_{\textrm{min}}$ and $\alpha_{\textrm{max}}$ represent the minimum and maximum achievable conductance values of the memristors, respectively. Let $\bf{O}$ denote the matrix to be mapped, and let $\bf{U}$ and $\bf{V}$ denote the two corresponding conductance matrices. When mapping $\bf{O}$ onto ${\bf{U}}-{\bf{V}}$, we consider the following mapping scheme:
\begin{equation}
  u_{i,j} = \begin{cases}
    {\alpha_{\textrm{max}},o_{i,j} > 0} \\
    {\alpha_{\textrm{min}},o_{i,j} \leq  0}
    \end{cases}
\end{equation}
and
\begin{equation}
  v_{i,j} = u_{i,j} - \beta o_{i,j},
\end{equation}
where $\beta$ is given by $\frac{\alpha_{\textrm{max}}-\alpha_{\textrm{min}}}{\textrm{max}\{\left|o_{i,j}\right|\}}$, ensuring that the conductance range covers all elements of $\bf{O}$. In this paper, we consider the memristor conductance range of $0.1\,\rm{\mu}$S $\sim 30\,\rm{\mu}$S.

\subsection{Detection Performance}\label{S4.1}

\begin{figure}[tbp]
  \centerline{\includegraphics[width=0.9\linewidth]{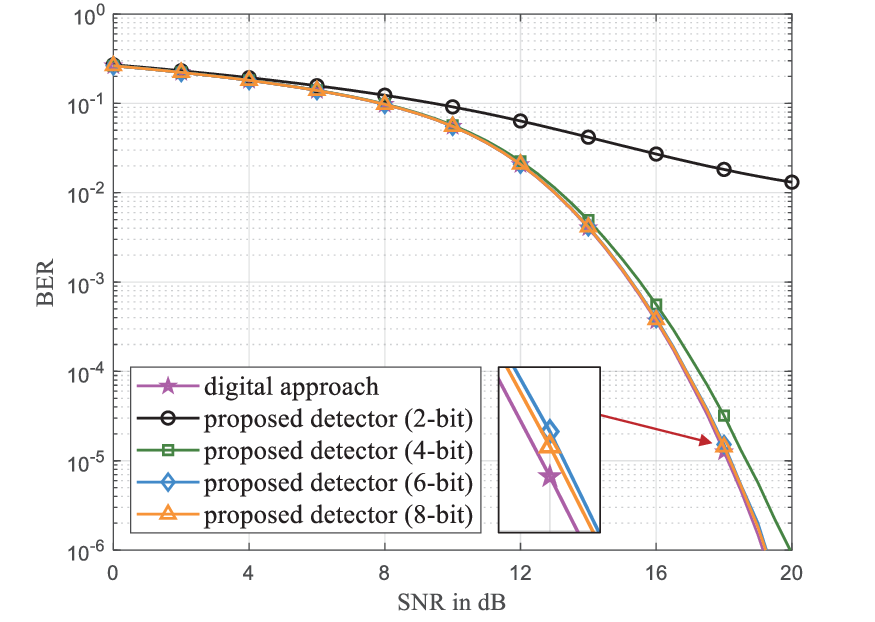}}
\vspace*{-4mm}
\caption{BERs of the proposed detector circuit under different values of memristor precision, using the digital approach as the benchmark.}
  \vspace*{-4mm}
\label{BER} 
\end{figure}

Unlike digital processors, the computational results of a memristor-based analog matrix computing circuit are not absolutely precise and are particularly affected by the precision of the memristors. In this experiment, we consider a $32\times 64$ massive MIMO system, i.e., $K=32$ and $R=64$, using 16 QAM. All UTs are assumed to have the same transmitted power and the column norm-based ordering scheme \cite{MIMOOFDM} is adopted. Fig.~\ref{BER} depicts the bit error rates (BERs) of the proposed detector circuit as the functions of the signal-to-noise ratio (SNR) under different memristor bit-precision values, using the digital approach as the benchmark. As expected, the higher the precision of the memristors, the lower the BER of the proposed detector. The simulation results indicate that the memristor bit-precision should be at least 6 bits to ensure detection performance approaching the BER of digital computing.

\subsection{Computing Time, Computing Speed and Computational Energy Efficiency}\label{S4.2}

We consider a $4\times 4$ MIMO system, i.e., $K=R=4$, in the noise-free environment to investigate the rules of the convergence time of the proposed detector circuit. OAs in the circuit have a gain-bandwidth product (GBP) of 500\,MHz, and $v_0$ is set to 0.1\,V. SPICE simulation results presented in this paper are provided by LTspice$^\circledR$.

\begin{figure}[bp]
  \vspace*{-5mm}
  \centering
    \subfloat[]{\includegraphics[width=0.92\linewidth]{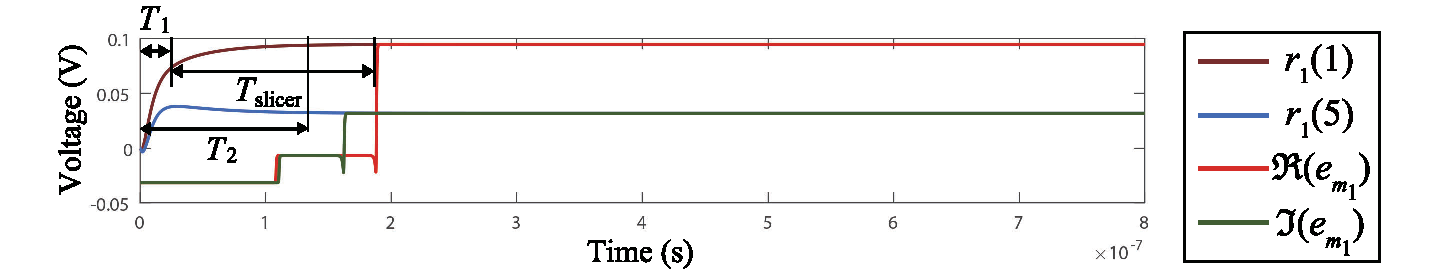}}\vspace*{-3mm}
    \hfil
    \subfloat[]{\includegraphics[width=0.92\linewidth]{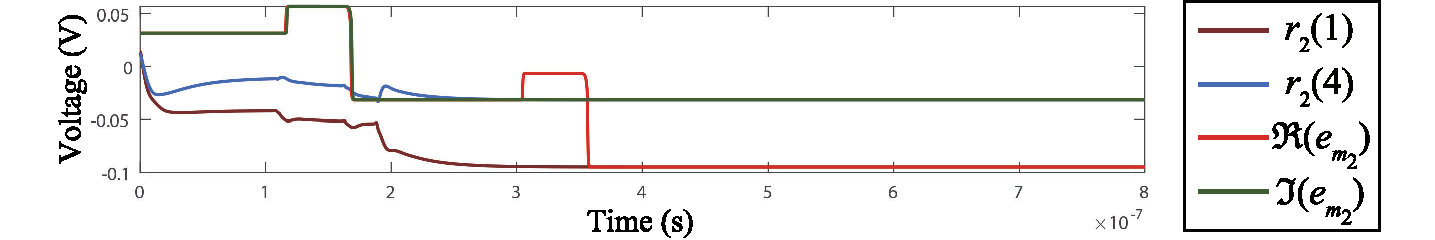}}\vspace*{-3mm}
    \hfil
    \subfloat[]{\includegraphics[width=0.92\linewidth]{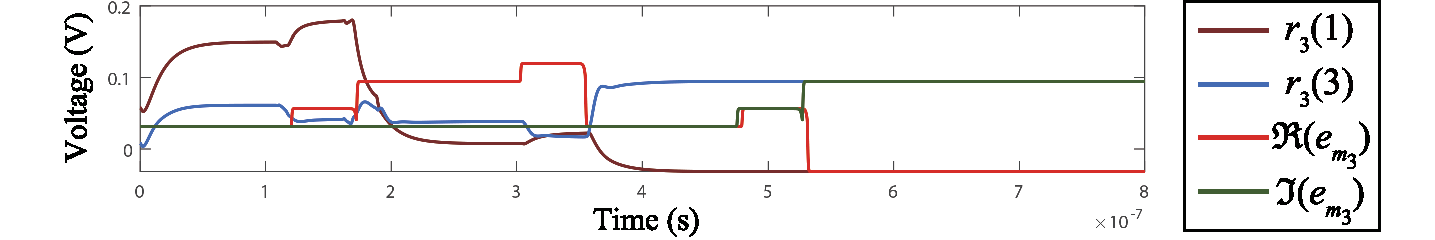}}\vspace*{-3mm}
    \hfil
    \subfloat[]{\includegraphics[width=0.92\linewidth]{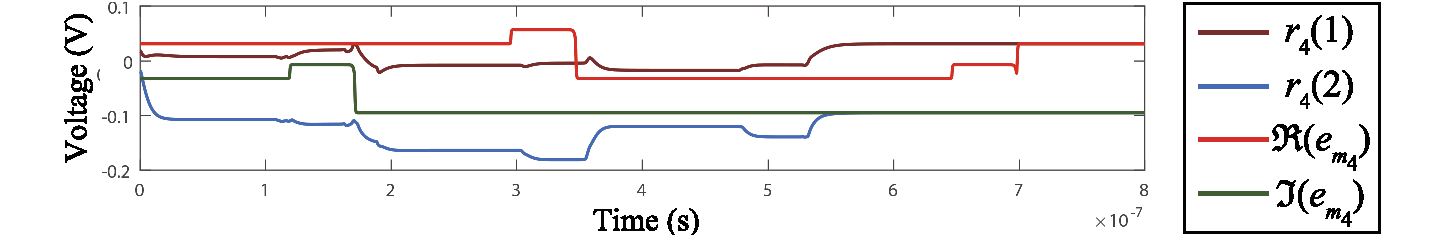}}\vspace*{-1.5mm}
    \caption{Voltage waveforms at input and output nodes of slicers in the four stages of the proposed circuit: (a) the first stage, (b) the second stage, (c) the third stage, (d) the fourth stage.}
  \vspace*{-0.5mm}
  \label{waveall} 
\end{figure}

We first consider the directly select structure, employing the voltage comparator LT1016 \cite{LT1016} and the 8-channel multiplexer ADG1608 \cite{ADG1608}. Let $T_{\textrm {slicer}}$ be the delay of the proposed slicer, i.e., the delay time between the input of the slicer reaching the threshold and the slicer completing the switching. $T_{\textrm {slicer}}$ is the sum of the propagation delay of the comparator, which is 10\,ns typically, and the transition time of the multiplexer, which is 150\,ns typically. Fig.~\ref{waveall} shows an example of voltage waveforms at input and output nodes of slicers in the four stages of the proposed detector circuit. For the first stage, the convergence time of $\Re(e_{m_1})$ is $T_{1}+T_{\textrm {slicer}}$, where $T_{1}$ is the time required for $r_1(1)$ to reach the corresponding threshold. Let $T_{2}$ be the time required for the outputs of a matrix computing module to reach steady states. Obviously, we have $T_{1}\leq T_{2}$, indicating that the maximum convergence time of the first stage in the circuit is $T_{2}+T_{\textrm {slicer}}$. The change in the output of a slicer leads to alterations in the inputs of subsequent slicers, which may result in changes in the outputs of these slicers. Therefore, the maximum convergence time of the proposed circuit is $K(T_{2}+T_{\textrm {slicer}})$, which corresponds to the longest time required for the output voltages of the $K$th stage to reach steady state after inputs are applied. As for the indirectly select structure, the delay of the combinational logic circuit is negligible, and therefore the convergence time of the circuit with indirectly select structure also satisfies the above rules.

\begin{figure}[tbp]
  \vspace*{-1mm}
  \centerline{\includegraphics[width=0.85\linewidth]{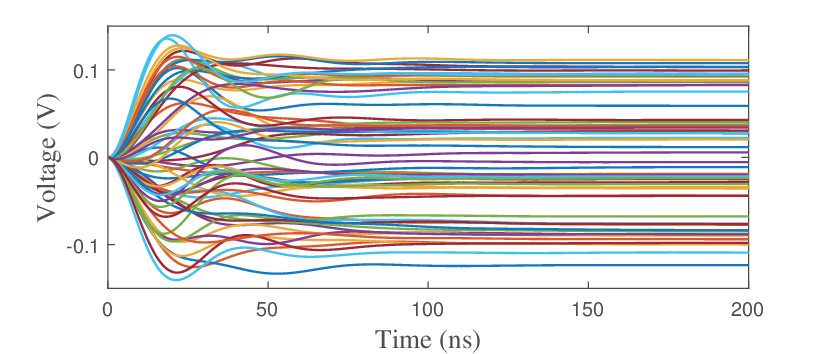}}
\vspace*{-4mm}
\caption{An example of the voltage waveforms of the 64 output nodes in the first matrix computing module.}
\vspace*{-4mm}
\label{MCMwave} 
\end{figure}

In the rest of this subsection, we consider a $32\times 64$ massive MIMO system. To further reduce the convergence time, we consider the high-speed voltage comparator MAX903 with 8\,ns propagation delay \cite{MAX903}, the 4-channel multiplexer ADG709 \cite{ADG708709}, and the 8-channel multiplexer ADG708 \cite{ADG708709}. Both the ADG708 and ADG709 exhibit a latency of 14\,ns. Fig.~\ref{MCMwave} shows an example of voltage waveforms of the 64 output nodes in the first matrix computing module, i.e., ${\mathbf r}_1$. $T_{2}$ is typically less than 130\,ns and can be further reduced by increasing the GBP of OAs \cite{RidgeRegression}. Since the considered 4-channel and 8-channel multiplexers have the same delay time, there is no difference in the convergence time of the detector circuits with the two slicer structures. Analog to digital converters (ADCs) of \cite{ADCref} with 10\,ns delay are used to measure the output voltages of the circuit. OAs, in conjunction with the digital to analog converters (DACs) of \cite{CDACref} whose settling time is 0.4\,ns, are employed to provide stable input voltages for the circuit. In the considered scenario, the total computing time of the proposed detector circuit is approximately 4.87\,$\rm{\mu}$s, including the settling time of DACs, the convergence time of the detector circuit, and the delay of ADCs. The computing speed of the proposed circuit is evaluated by the ratio of its equivalent floating-point operation (FLOP) number to its computing time. In the considered scenario, the computing speed of the proposed circuit is about 5.5 tera operations per second (TOPS).

To further evaluate the energy consumption of the proposed circuit, we consider the OA of \cite{OAref} with a GBP of 500\,MHz and power consumption of 12\,$\rm{\mu}$W. We consider both the Joule dissipation on each memristor and typical energy consumption of all other components in the circuit during the computing period. In the scenario considered, the energy consumption of the circuits with both the directly select structure and the indirectly select structure is approximately the same, at 18.98\,$\rm{\mu}$J. The computational energy efficiency of a circuit can be expressed as the ratio of its equivalent FLOP number to the energy consumed for completing a given computational task, and it is typically measured in TOPS per watt (TOPS/W), i.e., tera operations per joule. In the scenario considered, the proposed detector circuit achieves the computational energy efficiency of approximately 1.41 TOPS/W.

In Table~II we compare the proposed circuit with three different traditional digital computing approach based processors, including a DSP, an FPGA, and a GPU, in terms of equivalent computing speed, energy consumption, and computational energy efficiency. Specifically, we use Texas Instruments’ 8-core high-performance DSP TMS320C6678 \cite{DSPref}, the high-performance low-power FPGA Xilinx Virtex-7 690T \cite{FPGA}, and the powerful GPU NVIDIA RTX A1000 \cite{GPU} as representative models for these three types of processors, respectively.

The proposed circuit operates at a speed approximately 43 times faster than that of the DSP, with the computational energy efficiency around 110 times greater, i.e., the computing energy consumption is merely less than 1\% of that of the DSP. Compared with the Virtex-7 690T FPGA, the proposed circuit exhibits a computing speed and computational energy efficiency approximately 1.76 and 18 times greater, respectively. Although the equivalent computing speed of the proposed circuit is slightly lower than that of the RTX A1000 GPU in the considered scenario, its computational energy efficiency exceeds that of the RTX A1000 GPU by more than an order of magnitude. Furthermore, for massive MIMO systems with a higher number of UTs or BS antennas, the proposed memristor-based detector circuit can achieve even higher equivalent computing speed and computational energy efficiency \cite{RidgeRegression}, making it particularly suitable for next-generation wireless communication systems with massive connectivity requirements.

\section{Conclusion}\label{S5}

In this paper, we have presented a memristor-based massive MIMO SIC detector circuit for $K$ UTs. The proposed detector circuit structure comprises $K$ stages, each containing a matrix computing module and two slicers. We have explained the circuit structure of the matrix computing modules and proposed two feasible slicer structures. We have also investigated the impact of the precision of memristors on the detection performance of the proposed detector circuit. In a $32\times 64$ massive MIMO scenario, we have evaluated the computing speed and computational energy efficiency of the proposed circuit, demonstrating significant advantages over the representative high performance DSP, FPGA and GPU.

\renewcommand{\arraystretch}{1.45}
\begin{table*}[tbp]
\caption{Comparison between the proposed circuit and different traditional digital computing approach based processors}
\vspace*{-2mm}
\label{comparison} 
  \centering
  \begin{tabular}{>{\centering}p{0.25\linewidth}>{\centering}p{0.14\linewidth}>{\centering}p{0.14\linewidth}>{\centering}p{0.14\linewidth}p{0.14\linewidth}<{\centering}}
  \hline
  & \makecell{8-core DSP\\(TMS320C6678)} & \makecell{FPGA\\(Virtex-7 690T)} & \makecell{GPU\\(RTX A1000)} & {\bf Proposed circuit} \\
  \hline
  Computing speed & 0.128\,TOPS & \makecell{3.12\,TOPS} & \makecell{6.7\,TOPS} & {\bf 5.5\,TOPS}\\
  Energy consumption & 2.1\,mJ & \makecell{343.1$\mu$J} & \makecell{199.7$\mu$J} &{\bf 18.98\,$\bm{\mu}$J}\\
  Computational energy efficiency &0.0128\,TOPS/W & \makecell{0.078\,TOPS/W} & \makecell{0.134\,TOPS/W} &{\bf 1.41\,TOPS/W}\\
  \hline
  \end{tabular}
\vspace*{-2mm}
\end{table*}

\bibliographystyle{IEEEtran}

\end{document}